\documentclass[10pt,conference,letterpaper]{IEEEtran}
\IEEEoverridecommandlockouts
\usepackage{booktabs}
\usepackage{bm}
\usepackage{graphicx}
\usepackage[cmex10]{amsmath}
\usepackage{cite}
\usepackage{amssymb}
\usepackage{subfigure}
\usepackage{extarrows}
\usepackage[letterspace = -2]{microtype}
\usepackage{algorithm}
\usepackage{algpseudocode}
\usepackage{float}
\usepackage{color}
\usepackage{xcolor}
\usepackage{algorithmicx}
\usepackage{verbatim}
\newfloat{algorithm}{t}{lop}

\begin{document}

\title{ Physical-Layer Security for 6G: Safe Jamming against Malicious Sensing\\
\thanks{This work was partially supported by the National Natural Science Foundation of China under Grant U2001210, 62211540396, 61901216, and the Key R\&D Plan of Jiangsu Province under Grant BE2021013-4. (Corresponding author: Yang Huang)}
}
\author{
\IEEEauthorblockN{Pu Xie\IEEEauthorrefmark{1}, Yang Huang\IEEEauthorrefmark{1}}
\IEEEauthorblockA{\IEEEauthorrefmark{1} College of Electronic and Information Engineering, Nanjing University of Aeronautics and Astronautics}
\IEEEauthorblockA{Email: yang.huang.ceie@nuaa.edu.cn}
}

\maketitle

\vspace{-0.2cm}
\begin{abstract}
The integration of sensing, communications, array signal processing, etc. into 6G mobile networks has ushered in an era of heightened situational awareness. However, this progress brings forth significant concerns regarding privacy and security, particularly due to the proliferation of devices equipped with radar-like sensing capability, including malicious ones. In response, this paper proposes a novel actor-critic (AC) method-based frequency selection scheme for noise jamming, in order to effectively counter malicious multifunction frequency agility sensing. In the meanwhile, to mitigate potential interference (caused by sidelobes of the jamming beam) with uplink transmissions conducted by legitimate but non-cooperative users, a robust action correction mechanism, which is capable of learning and predicting the spectrum utilization state, is proposed to find feasible but near-optimal frequency configuration for jamming. Numerical results demonstrate that benefiting from the robust action correction mechanism, the proposed AC-based safe jamming can not only make the malicious sensing device continuously get stuck in the searching mode but also guarantee minimal disruption to the legitimate non-cooperative users.
\end{abstract}

\begin{IEEEkeywords}
6G, safe jamming, robust action correction, actor-critic.
\end{IEEEkeywords}

\section{Introduction}
\label{INTRODUCTION}
Sensing is one of the main driving forces of technologies forming the 6G mobile communication system \cite{MAHG23}. In the 6G mobile communication system, devices can act as sensors, exploiting the transmission, reflection, and scattering of radio waves to sense the physical world. Distance, speed, and angle information can be derived from wireless signals to to lay the foundations of detecting and tracking passive objects, imaging, and other sensing capabilities. Malicious sensing device can locate, track, illegally acquiring information on target activity and determining the nature of target terrain, forming a threat to public security. However, traditional physical layer security methods \cite{QQ23} such as secure coding only ensure communication security through pre-coding signal processing, while inapplicable to guarantee the physical security of 6G system, in terms of preventing illegal sensing.

To effectively interfere with malicious sensing devices and disrupt their sensing capabilities, a series of jamming methods \cite{BKBYYH23} \cite{ZYYGWG22} \cite{GYWJYH23} have been investigated to address multifunction and frequency agility malicious sensing devices.
Considering that frequency-agile malicious sensing devices can resist interference by changing their frequencies, a smart jamming strategy design method and periodic action evaluation decision-making method based on reinforcement learning (RL) were proposed in \cite{BKBYYH23}. For multifunction devices, \cite{GYWJYH23} proposed a cognitive jamming method based on threat assessment, where the jamming decision-making problem was modeled as a markov decision process and rewards were formulated based on the trajectory-based threat assessment models. Ref. \cite{ZYSY23} proposed a jamming scheme against multifunction device based on actor-critic (AC) method. It is worth noting that the aforementioned studies only focus on interfering with multifunction or frequency agility malicious sensing devices, while neglect the urgent demand of avoiding interference (caused by the sidelobes of the jamming beam) with nearby friendly communication and/or sensing devices.

The above issue of dynamically configuring wireless resources for safe jamming, i.e. jamming malicious sensing devices while avoid interference with friendly communication or sensing devices, always boils down to a sequential decision making problem, which is usually solved by RL, with constraints. In terms of solving such constrained decision-making problems, \cite{YLSTJ23} \cite{MSDAA23} \cite{N23} investigated RL suffering from constraints. An online frequency agility strategy based on a multi-armed bandit model is proposed in \cite{YLSTJ23}, optimizing detection performance under dynamic frequency response without causing interference. Additionally, a method utilizing online RL and multiple Q-learning algorithms is introduced in \cite{MSDAA23} to optimize drone trajectory planning under the constraints of dynamically changing obstacle. Ref. \cite{N23} proposed an overall framework for vehicle network routing based on RL, where peak constraints on the maximum delay or minimum bandwidth and long-term average constraints on the average transmit power or data rate can be simultaneously satisfied. Nevertheless, methods proposed in the aforementioned studies are inapplicable to the scenario where constraints needs to be strictly satisfied in each timeslot where actions selected in the previous timeslot are executed.

\begin{figure}[t]
\centering
\includegraphics[width = 3.4in]{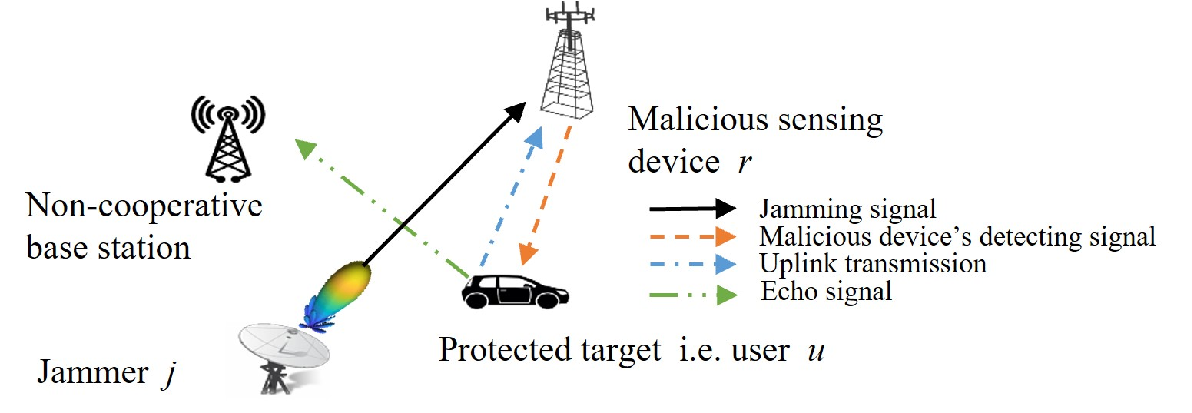}
\caption{Safe jamming in 6G system.}
\label{FigSystemModel}
\end{figure}
In this paper, we investigate the scenario of safe jamming in the 6G system, as shown in Fig. \ref{FigSystemModel}. Basically, this paper focuses on addressing the problem of dynamically configuring frequency-domain resources, so as to effectively jam the malicious sensing device but not interfere with friendly and non-cooperative uplink transmissions (due to the sidelobe of the jamming beam).
Nevertheless, solving the constrained frequency configuration policy optimization for jamming malicious sensing device but strictly avoid interfering with non-cooperative uplink transmissions in each timeslot is non-trivial.
Therefore, we firstly propose a novel AC method-based frequency selection algorithm for noise jamming without the hard constraints (which takes effect in the next timeslot) on frequency conflict.
In order to tackle the issue of confirming whether the frequency selected in the current timeslot violates the constraints that become effective in the future timeslot, approximations of the constraint functions are trained to predict the spectrum occupation situation.
Moreover, in the presence of potential future frequency-domain conflicts, a near-optimal frequency configuration can be obtained with the robust action correction scheme, where closed-form solution is obtained by analyzing the Karush-
Kuhn-Tucker (KKT) conditions.
Simulation results show that in the offline training phase, the proposed AC method-based algorithm with robust action correction can converge slower than that without robust action correction, due to the action modification processes for safe jamming in the training phase.
Despite this, the proposed AC-based safe jamming can keep the malicious sensing device always operating in the searching mode and simultaneously avoid interference with the friendly but non-cooperative users.

The remainder of this paper is organized as follows. Section \ref{SecSystemModel} elaborates on the system model. Section \ref{SecProbFormulation} formulates the design problem. Section \ref{ACMBSJ} proposes the AC method-based algorithm with robust action correction. Section \ref{SecSimResults} discusses the simulation results. Conclusions are drawn in Section \ref{SecConclusions}.

\section{System Model}
\label{SecSystemModel}
As shown in Fig. \ref{FigSystemModel}, this paper focuses on the scenario of a typical 6G mobile network, where the entire frequency band consists of $N$ frequency domain channels.
In the studied scenario, a malicious sensing device $r$, which can be a fake base station, continuously transmits a radar-like sensing signal to detect user $u$.
Afterwards, target information can be extracted from echo signals by the sensing device for target detection and recognition. In order to protect the user from the malicious detection, a jammer $j$ is deployed to monitor the spectrum usage and interfere with the malicious sensing by noise jamming.
We consider that in a certain timeslot $t$, without loss of generality, if the jamming signal and the sensing signal occupy the same frequency, jamming is supposed to be successful, assuming that the transmit power of the jammer is adequately high.
In the meanwhile, we consider that a user $u$ performs uplink transmission with frequency hopped waveforms and sends data to a non-cooperative base station, operating at the same frequency band as sensing.
The terminology ``non-cooperative" means that the jammer and the base station cannot be coordinated to share the frequency band and the dynamics of frequency hopping for uplink transmissions is unknown to the jammer.

\begin{figure}[t]
\centering
\includegraphics[width = 3.4in]{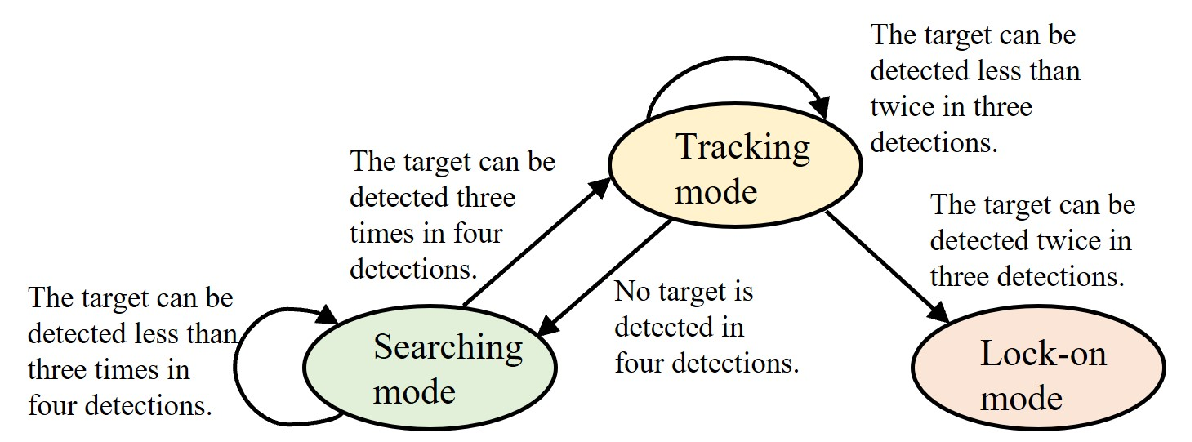}
\caption{Transition of the modes during sensing.}
\label{Figstate}
\end{figure}
We consider a frequency-agile multifunction malicious sensing device \cite{ZYSY23}. It means that in order to mitigate the effect of noise jamming at the receiver of the sensing device $r$, the transmitter of the sensing device employs a predefined frequency hopping schedule to adjust the carrier frequency for transmitting the radar-like sensing signals over time.
Moreover, such a multifunction sensing device can switch the modes of operation according to the state transition diagram, as depicted in Fig. \ref{Figstate}.
Given that the multifunction sensing device operates in the searching mode, if a target can be detected three times in four detections, the sensing device switches to the tracking mode; otherwise, the sensing device continuously gets stuck in the searching mode.
When operating in the tracking mode, if the target can be detected twice in three detections, the sensing device switches to the lock-on mode; however, if the target can be detected less than twice in three detections but at least one time in four detections, the malicious sensing device continuously operating in the tracking mode; otherwise, the sensing device switches back to the searching mode.
Once the malicious sensing device switching to the lock-on mode, it means the failure of jamming, as well as the terminal of the confrontation.

In the subsequent discussions, $a_{j,t,n}=1$ means that in timeslot $t$, jammer $j$ selects frequency $n \in \{1, \ldots, N\}$ (to be utilized in timeslot $t+1$); otherwise, $a_{j,t,n} = 0$. Moreover, if user $u$ utilizes frequency $n$ for the uplink transmission in timeslot $t+1$, $y_{u,t,n} = 1$; otherwise, $y_{u,t,n} = 0$. Hence, in timeslot $t+1$, for no spectrum conflicts between jammer $j$ and user $u$, it has to be satisfied that
\begin{equation} \label{constraint}
\begin{split}
a_{j,t,n}+y_{u,t+1,n}\le 1,\forall n.
\end{split}
\end{equation}

Jammer $j$ is equipped with a reconnaissance module, which is capable of not only recognizing the uplink waveforms transmitted by user $u$ (through detecting the preambles in the waveforms) but also measuring the received signal strength (RSS) across frequencies for the entire frequency band.
Note that if only the measured RSS at the frequency occupied by the noise jamming is significantly higher than a predefined threshold (which is used to distinguish a valid signal from noise), it means that the noise jamming signal coincides with the sensing signal delivered by the malicious sensing device in frequency domain.
Hence, in timeslot $t$, jammer $j$ can observe the frequency utilization $f_{r,t} \in \{1, \ldots, N\}$ of malicious sensing device $r$, as well as the frequency utilization $f_{u,t} \in \{1, \ldots, N\}$ of user $u$.
Moreover, the reconnaissance module is capable of identifying the modes of operation $m_{r,t}$ of the malicious sensing device. Remarkably, the transition of the modes can be exploited to estimate whether the malicious sensing device is jammed.

\section{Problem Formulation}
\label{SecProbFormulation}
This paper addresses the problem of safe jamming, where the jammer is required to always interfere with the malicious sensing device (such that it gets stuck in the searching mode) without affecting the user's uplink transmissions.
Nevertheless, according to Section \ref{SecSystemModel}, in a certain timeslot $t$, the jammer has no knowledge of the frequency occupation $f_{r,t+1}$ or $f_{u,t+1}$. Intuitively, in order to determine frequency $a_{j,t,n}$ for jamming in timeslot $t+1$, the jammer has to rely on the observation of the spectrum utilization \cite{YCYF21} made in timeslot $t$. Therefore, the design problem boils down to a Markov decision process (MDP) \cite{RA18}.

In order to formulate the MDP, we designate the state space $\mathbf{S}$ as a set that collects states of the spectrum utilization in a timeslot, while the action space $\mathbf{A}$ is a set that collects all frequency-domain resources for jamming in a timeslot.
According to Section \ref{SecSystemModel}, in timeslot $t$, the spectrum utilization state ${\mathbf{s}_{j,t}}\in \mathbf{S}$ observed by jammer $j$ can be fomulated as ${\mathbf{s}_{j,t}}=\left[ \mathbf{a}_{j,t-1}^T, f_{r,t}, f_{u,t}, m_{r,t} \right]^{T}$, where $\mathbf{a}_{j,t-1} = \left[ a_{j,t-1,1}, \ldots, a_{j,t-1,n}, \ldots , a_{j,t-1,N} \right]^T \in \mathbf{A}$ (for $\sum_{n=1}^{N} a_{j,t-1,n} =1$) means the frequency selected in timeslot $t-1$ for emitting noise jamming in timeslot $t$.

An immediate reward $r_{j,t}$ can be obtained by jammer $j$, indicating that the malicious sensing device is exactly interfered by the jammer, while the values of $r_{j,t}$ are different in various cases. If the noise jamming signal coincides with the sensing signal delivered by the malicious sensing device in frequency domain $r_{j,t} = R_{1}$; otherwise, $r_{j,t} = R_{2}$. If the mode of operation of the malicious sensing device is switched from the searching mode to the tracking mode (or from the tracking mode to the lock-on mode), it means the failure of jamming, and $r_{j,t} = R_{3}$. On the contrary, if the mode of operation of the malicious sensing device switches from the tracking mode to the searching mode, it means that the malicious sensing device is successfully jammed, and $r_{j,t} = R_{4}$.

Due to the unknown dynamics of frequency agility of the malicious sensing, as well as the unknown dynamics of frequency hopping for the non-cooperative uplink transmission, the transition probability $P\left( {\mathbf{s}_{j,t+1}}=\mathbf{s}'|{\mathbf{s}_{j,t}}=\mathbf{s},{\mathbf{a}_{j,t}}=\mathbf{a} \right)$, for $\mathbf{s}\in \mathbf{S}$ and $\mathbf{a}\in \mathbf{A}$, is unknown to the jammer. Due to this, intuitively, the formulated MDP has to be solved by RL\cite{RA18}.

Aiming at improving the expected long-term success rate of jamming and mitigating the interference with the friendly uplink transmission, the design problem of optimizing the frequency configuration policy for jamming can be cast as
\begin{subequations}
	\label{RLmain}
	\begin{IEEEeqnarray}{rl}
		\IEEEyesnumber
		\underset{{{\pi }_{j}}}{\mathop{\max }}\quad & E\left[ \sum\limits_{t=1}^{\infty }{{{\gamma }^{t-1}}{{r}_{j,t}}} \right]  \label{eq:sub1}\\
		s.t. \quad & b_n\left( {\mathbf{s}_{j,t}},{\mathbf{a}_{j,t}} \right) \triangleq a_{j,t,n}+y_{u,t+1,n} \leq 1,\, \forall n \label{eq:sub2} \\
		& a_{j,t,n}\in \left\{ 0,1 \right\} \label{eq:sub3}
	\end{IEEEeqnarray}
\end{subequations}
where ${{\pi }_{j}}:\mathbf{S}\to \mathbf{A}$ represents a deterministic policy w.r.t. the frequency selection at jammer $j$; $\gamma \in \left( 0,1 \right)$ represents a discount factor.

\section{Actor-Critic Method-Based Safe Jamming}
\label{ACMBSJ}
This section proposes an RL algorithm with robust action correction to dynamically configure frequency resources for safe jamming.

\subsection{Jamming Strategy Optimization with AC-Based Method}
\label{SubSecAC}
Solving the constrained policy optimization problem (\ref{RLmain}) is non-trivial. Therefore we firstly solve the problem without constraints. Conventional methods based on value function approximation such as deep Q-network may not converge to any stable strategy due to discontinuous variations in the estimated value of an action, while conventional policy gradient methods may not update policy parameters in a direction of improvement in  the expected discounted reward \cite{YXQ23}. Therefore in order to avoid the aforementioned issue, we employ an AC approach to solve the jamming strategy optimization, i.e. \eqref{eq:sub1}. Briefly, the AC approach alternately improves an estimate of the state-value function and the parameterized policy ${\pi}_{j} \left( \cdot \left| {\mathbf{s}_{j,t}};{{\theta }_{j}} \right. \right)$, by learning the critic parameters ${{w}_{j}}$ and the actor parameters ${{\theta }_{j}}$.
The actor generates frequency selection according to the spectrum utilization state, by exploiting a parameterized policy ${\pi}_{j} \left( \cdot \left| {\mathbf{s}_{j,t}};{{\theta }_{j}} \right. \right)$. The critic assesses the action taken by the actor with a parameterized approximate state-value function $V^{{{\pi }_{j}}}\left( {\mathbf{s}_{j,t}};{{w}_{j}} \right)$ and exploits the advantage to update the policy network parameters ${{\theta }_{j}}$ and the network parameters ${{w}_{j}}$.

In order to evaluate the optimality of executing a frequency selection action ${\mathbf{a}_{j,t}}$ with a given spectrum utilization state ${\mathbf{s}_{j,t}}$, the AC method establish an advantage function. The advantage contributes to the maximization of the state-value function and can be formulated as
\begin{equation}
\begin{split}
{{A}^{{{\pi }_{j}}}}\left( {\mathbf{s}_{j,t}},{\mathbf{a}_{j,t}} \right)={{Q}^{{{\pi }_{j}}}}\left( {\mathbf{s}_{j,t}},{\mathbf{a}_{j,t}} \right)-{{V}^{{{\pi }_{j}}}}\left( {\mathbf{s}_{j,t}} \right).
\end{split}
\end{equation}
However, due to the unknown ${{Q}^{{{\pi }_{j}}}}\left( {\mathbf{s}_{j,t}},{\mathbf{a}_{j,t}} \right)$ and ${{V}^{{{\pi }_{j}}}}\left( {\mathbf{s}_{j,t}} \right)$, the aforementioned advantage cannot be directly obtained. Hence, the advantage can be approximated as
\begin{equation}
\begin{split} \label{advantage}
{{A}^{{{\pi }_{j}}}}\left( {\mathbf{s}_{j,t}},{\mathbf{a}_{j,t}} \right)\approx {{r}_{j,t}}+{{V}^{{{\pi }_{j}}}}\left( {\mathbf{s}_{j,t+1}} \right)-{{V}^{{{\pi }_{j}}}}\left( {\mathbf{s}_{j,t}} \right).
\end{split}
\end{equation}

The gradient with respect to (w.r.t.) the state value in actor-critic algorithm can be cast as
\begin{equation}
\begin{split}
\nabla J\left( {{\theta }_{j}} \right)=\sum\limits_{t=1}^{\tau }{\nabla \log {{\pi }_{j}}\left( {\mathbf{s}_{j,t}};{{\theta }_{j}} \right){{A}^{{{\pi }_{j}}}}\left( {\mathbf{s}_{j,t}},{\mathbf{a}_{j,t}} \right)}.
\end{split}
\end{equation}
In order to maximize the advantage i.e. (\ref{advantage}), the critic network parameters can be updated as
\begin{equation}
\begin{split}
{{w}_{j}}\leftarrow {{w}_{j}}+\nabla {\left[ A^{{{\pi }_{j}}}\left( {\mathbf{s}_{j,t}},{\mathbf{a}_{j,t}} \right) \right]^{2}}.
\end{split}
\end{equation}
In order to maximize $J$, the actor network parameters can be updated by
\begin{equation}
\begin{split}
{{\mathbf{\theta }}_{j}}\leftarrow {{\mathbf{\theta }}_{j}}\text{+}\alpha \cdot \nabla J({{\mathbf{\theta }}_{j}}).
\end{split}
\end{equation}

\subsection{Robust Action Correction Scheme}
\label{SubSecRobust}
Although subsection \ref{SubSecAC} solves the dynamic frequency-domain resource configuration problem for jamming without any constraint, the jammer needs to ensure that no conflict occur in each timeslot, due to the constraint (\ref{eq:sub2}). To this end, a robust action correction scheme is then developed. Briefly, an approximation of the constraint function $b_n \left( {\mathbf{s}_{j,t}},{\mathbf{a}_{j,t}} \right)$ of spectrum utilization state is firstly constructed and exploited to predict the existence of frequency conflicts between the jammer and the friendly user. Then, in the presence of potential conflicts, a near-optimal frequency selection action can be computed to replace the action generated by the AC-based method.

Specifically, the jamming frequency $\mathbf{a}_{j,t}$ selected in timeslot $t$ can only be executed in the future timeslot $t+1$. Due to the unknown transition probability $P\left( {\mathbf{s}_{j,t+1}} | {\mathbf{s}_{j,t}},{\mathbf{a}_{j,t}} \right)$, the value of the constraint function $b_n\left( {\mathbf{s}_{j,t}},{\mathbf{a}_{j,t}} \right)$ can only be observed in the future timeslot $t+1$. Therefore, in order to predict whether $\mathbf{a}_{j,t}$ causes frequency conflict in timeslot $t+1$, we need an approximation of $b_n \left( {\mathbf{s}_{j,t}},{\mathbf{a}_{j,t}} \right)$, which can be observed in timeslot $t$. Hence, let $b_n\left( {\mathbf{s}_{j,t}},{\mathbf{a}_{j,t}} \right) \triangleq \overline{b_n}\left( {\mathbf{s}_{j,t+1}} \right)$, a parameterized linear approximation of $\overline{b_n}\left( {\mathbf{s}_{j,t+1}} \right)$ can be constructed as
\begin{equation} \label{8}
\begin{split}
\overline{b_n}\left( {\mathbf{s}_{j,t+1}} \right)\approx \overline{b_n}\left( {\mathbf{s}_{j,t}} \right)+k_n{{\left( {\mathbf{s}_{j,t}},w_n \right)}} {{a}_{j,t,n}}, \, \forall n,
\end{split}
\end{equation}
where $w_n$ represent the parameters of the neural network characterizing $k_n{{\left( {\mathbf{s}_{j,t}},w_n \right)}}$.
Then, (\ref{8}) can be recast as
\begin{equation}
\begin{split}
b_n\left( {\mathbf{s}_{j,t}},{\mathbf{a}_{j,t}} \right)\approx \overline{b_n}\left( {\mathbf{s}_{j,t}} \right)+k_n{{\left( {\mathbf{s}_{j,t}},w_n \right)}} {{a}_{j,t,n}}, \, \forall n,
\end{split}
\end{equation}
such that the constraint (\ref{eq:sub2}) can be rewritten as
\begin{equation}
\begin{split}
\overline{b_n}\left( {\mathbf{s}_{j,t}} \right)+k_n{{\left( {\mathbf{s}_{j,t}},w_n \right)}} {{a}_{j,t,n}}\le 1, \, \forall n,
\end{split}
\end{equation}

In order to learn the values of $w_n$ for $\overline{b_n}\left( \cdot \right)$, in the offline training phase, after executing the frequency selection action of a timeslot, the spectrum utilization state ${\mathbf{s}_{j,t}}$, the frequency selection action ${\mathbf{a}_{j,t}}$ and the state of next timeslot ${\mathbf{s}_{j,t+1}}$ can be obtained.
Hence, the values of $\overline{b_n}\left( {\mathbf{s}_{j,t}} \right)$, $k_n\left(\mathbf{s}_{j,t},w_n \right) a_{j,t,n}$ and $b_n\left( {\mathbf{s}_{j,t+1}} \right)$ can be computed. Aiming at minimizing the gap between the linearized approximation (\ref{8}) and the original $\overline{b_n}\left( {\mathbf{s}_{j,t+1}} \right)$, the parameter $w^\ast_n$ can be obtained by solving the following optimization \cite{GKMTCY18}:
\begin{equation}
\begin{split}
\textstyle{\min_w} \left| \overline{b_n}\left( {\mathbf{s}_{j,t}} \right) + k_n\left(\mathbf{s}_{j,t},w_n \right) a_{j,t,n} - \overline{b_n}\left( {\mathbf{s}_{j,t+1}} \right) \right|^2 \, .
\end{split}
\end{equation}

In the online inference phase, based on the spectrum utilization state ${\mathbf{s}_{j,t}}$ and the frequency selection action ${\mathbf{a}_{j,t}}$ output from the AC method-based frequency selection algorithm, the value of the constraint function $\overline{b_n}\left( {\mathbf{s}_{j,t}} \right)+k_n{{\left( {\mathbf{s}_{j,t}},w_n \right)}} {{a}_{j,t,n}}$ can be obtained, given the frequency selection action ${\mathbf{a}_{j,t}}$ and spectrum utilization state ${\mathbf{s}_{j,t}}$.
If $\overline{b_n}\left( {\mathbf{s}_{j,t}} \right)+k_n{{\left( {\mathbf{s}_{j,t}},w_n \right)}}\cdot{{a}_{j,t,n}}\le 1$, it means that the frequency selection action ${\mathbf{a}_{j,t}}$ yield by the AC method-based frequency selection algorithm would not conflict with the frequency utilization for user $u$'s uplink transmission in the future timeslot, and it is unnecessary to replace ${\mathbf{a}_{j,t}}$.
On the contrary, in the presence of $\overline{b_n}\left( {\mathbf{s}_{j,t}} \right)+k_n{{\left( {\mathbf{s}_{j,t}},w_n \right)}} {{a}_{j,t,n}}>1$, it means that the frequency configuration for noise jamming obtained by the AC method-based frequency selection algorithm may conflict with the user's communication frequency in the next timeslot, and it is necessary to replace the original ${\mathbf{a}_{j,t}}$ with a near-optimal but safe jamming frequency.

In order to make the corrected action not violate the constraints (\ref{eq:sub2}) and close to the action achieved by performing the strategy that only maximizes the long-term average reward (\ref{eq:sub1}) \cite{GKMTCY18}, the optimization of action correction can be formulated as
\begin{subequations}
	\label{main}
	\begin{IEEEeqnarray}{rl}
		\IEEEyesnumber
		\underset{\mathbf{{a}}_{j,c,t}}{\mathop{\min }}\quad & \frac{1}{2}{\left\| \mathbf{a}_{j,c,t}-{\mathbf{a}_{j,t}} \right\|}^{2}  \label{sub1}\\
		s.t. \quad & \overline{b_n}\left( {\mathbf{s}_{j,t}} \right)+k_n{{\left( {\mathbf{s}_{j,t}},w_n \right)}} {{a}_{j,c,t,n}}\le 1, \, \forall n \label{sub2} \\
		& a_{j,c,t,n}\in \left\{ 0,1 \right\} , \, \forall n \label{sub3}
	\end{IEEEeqnarray}
\end{subequations}
where $a_{j,c,t,n}$ is the $n$\,th element of the corrected action $\mathbf{a}_{j,c,t}$.
Due to (\ref{sub3}), problem (\ref{main}) turns out to be an integer programming problem, which is highly nonconvex.
Intuitively, (\ref{sub3}) can be relaxed, such that the constraint becomes $0 \leq a_{j,c,t,n} \leq 1$. Although such a relaxation can make the problem convex, it makes solving the problem slightly complex, introducing more Lagrangian multipliers.
Fortunately, given (\ref{8}), $b_n\left( {\mathbf{s}_{j,t}},{\mathbf{a}_{j,t}} \right) \triangleq \overline{b_n}\left( {\mathbf{s}_{j,t+1}} \right)$ and the physical meaning of $b_n\left( {\mathbf{s}_{j,t}},{\mathbf{a}_{j,t}} \right)$, it can be inferred that the value of $\overline{b_n}\left( {\mathbf{s}_{j,t}} \right)+k_n{{\left( {\mathbf{s}_{j,t}},w_n \right)}} {{a}_{j,c,t,n}}$ is roughly within the range of [0, 2], where the maximum value can be achieved only in the presence of $a_{j,c,t,n} = 1$.
This means that given $\mathbf{s}_{j,t}$ and $w$, $\overline{b_n}\left( \mathbf{s}_{j,t} \right)+k_n{{\left( {\mathbf{s}_{j,t}},w_n \right)}} {{a}_{j,c,t,n}}$ is monotonically increasing, such that satisfying (\ref{sub2}) can roughly guarantee that $0 \leq a_{j,c,t,n} \leq 1$.
Hence, (\ref{sub3}) can be omitted from (\ref{main}), yielding a convex problem, i.e.
\begin{subequations}
	\label{var_main}
	\begin{IEEEeqnarray}{rl}
		\IEEEyesnumber
		\underset{\tilde{\mathbf{a}}_{j,c,t}}{\mathop{\min }}\quad & \frac{1}{2}\left\| \tilde{\mathbf{a}}_{j,c,t}-\mathbf{a}_{j,t} \right\|^2 \label{var_sub1}\\
		s.t. \quad & \overline{b_n}\left( {\mathbf{s}_{j,t}} \right)+ k_n\left( \mathbf{s}_{j,t}, w_n \right) \tilde{a}_{j,c,t,n} \leq 1\, , \, \forall n\,. \label{var_sub2}
	\end{IEEEeqnarray}
\end{subequations}
The Lagrangian of problem (\ref{var_main}) can be obtained as
\begin{IEEEeqnarray}{l}
L\left( \tilde{\mathbf{a}}_{j,c,t}, \mathbf{\lambda}  \right) =  {\left\| \tilde{\mathbf{a}}_{j,c,t} - {\mathbf{a}_{j,t}} \right\|}^{2}/2 + \nonumber \\
\quad\quad \textstyle{\sum_{n=1}^N}\lambda_n \left( \overline{b_n}\left( {\mathbf{s}_{j,t}} \right)+k_n{{\left( {\mathbf{s}_{j,t}},w_n \right)}} {\tilde{a}_{j,c,t,n}} - 1 \right) \, ,
\end{IEEEeqnarray}
where $\mathbf{\lambda} = [\lambda_1, \ldots, \lambda_N]$ collects the Lagrangian multipliers w.r.t. the constraints (\ref{var_sub2}).
According to the KKT conditions, it can be obtained that
\begin{equation}
\overline{b_n}\left( {\mathbf{s}_{j,t}} \right)+ k_n\left( \mathbf{s}_{j,t}, w_n \right) \tilde{a}_{j,c,t,n}^\ast \leq 1\, ,\, \forall n
\end{equation}
\begin{equation}
\lambda_n^\ast \geq 0\, ,\, \forall n
\end{equation}
\begin{equation}
\nabla_{\tilde{a}_{j,c,t,n}}L= \tilde{a}^\ast_{j,c,t,n} - a_{j,t,n} + \lambda_n^\ast k_n\left( {\mathbf{s}_{j,t}}, w \right)=0\, ,\, \forall n
\end{equation}
\begin{equation}
\lambda_n^\ast \left( \overline{b_n}\left( {\mathbf{s}_{j,t}} \right)+k_n\left( \mathbf{s}_{j,t}, w_n\right) \tilde{a}_{j,c,t,n}^\ast - 1 \right )= 0\, ,\, \forall n
\end{equation}
By solving the above equations, the optimal $\tilde{a}^\ast_{j,c,t,n}$ can be achieved in closed-form, i.e.
\begin{equation}\label{tilde_ajctn}
\tilde{a}^\ast_{j,c,t,n} = a_{j,t,n} - \lambda_n^\ast k_n\left( {\mathbf{s}_{j,t}}, w \right) \, , \, \forall n.
\end{equation}
where $\lambda_n^\ast$ can be obtained as
\begin{equation}
\begin{split}
\lambda_n^{*}= \left[ \frac{k_n\left( \mathbf{s}_{j,t}, w_n\right) a_{j,t,n} + \overline{b_n}\left( {\mathbf{s}_{j,t}} \right) - 1 }{k_n^2\left( \mathbf{s}_{j,t}, w_n\right)} \right]^+ \,,\,\forall n \,.
\end{split}
\end{equation}
Since $\tilde{\mathbf{a}}^\ast_{j,c,t}$ is a continuous variable, it needs to be projected into a specific frequency selection action, within the original feasible domain of $\mathbf{a}_{j,c,t}$, yielding the optimized ${\mathbf{a}^\ast_{j,c,t}}$. The pseudocodes of the proposed RL with robust action correction are summarized in Algorithm \ref{ACbased}.

\begin{algorithm}[!h]
	\small{
		\caption{RL with robust action correction}\label{ACbased}
		\begin{algorithmic}[1]
			\State  \textbf{Initialize:} The state $\mathbf{s}_{j,0}$ of the jammer, the strategy network parameter ${{\theta }_{j}}$, and the value network parameter ${{w}_{j}}$.
			\Repeat
			\State In timeslot $t$, the jammer makes an observation of $\mathbf{s}_{j,t}$ and generates a frequency configuration by performing $\mathbf{a}_{j,t}={\pi}_{j} \left( \cdot |\mathbf{s}_{j,t};{{\theta }_{j}} \right)$ for emitting noise jamming in the future timeslot $t+1$;
			\State Compute $\overline{b_n}\left( {\mathbf{s}_{j,t}} \right)$ and $k_n{{\left( {\mathbf{s}_{j,t}},w_n \right)}} {{a}_{j,t,n}}$, with given $\mathbf{s}_{j,t}$ and $\mathbf{a}_{j,t}$ in the timeslot $t$;
			\If {$\overline{b_n}\left( {\mathbf{s}_{j,t}} \right)+k_n{{\left( {\mathbf{s}_{j,t}},w_n \right)}} {{a}_{j,t,n}}>1$}
			\State Obtain $\tilde{\mathbf{a}}^\ast_{j,c,t}$ by performing (\ref{tilde_ajctn});
            \State Project $\tilde{\mathbf{a}}^\ast_{j,c,t}$ to the original feasible domain of $\mathbf{a}_{j,c,t}$, yielding the optimized ${\mathbf{a}^\ast_{j,c,t}}$.
            \State Execute ${\mathbf{a}^\ast_{j,c,t}}$ in timeslot $t+1$;
			\ElsIf {$\overline{b_n}\left( {\mathbf{s}_{j,t}} \right)+k_n{{\left( {\mathbf{s}_{j,t}},w_n \right)}}\cdot{{a}_{j,t,n}}\le 1$}
			\State  Execute the original $\mathbf{a}_{j,t}$ in timeslot $t+1$;
			\State  By executing the frequency configuration in timeslot $t+1$, the jammer obtains the reward ${{r}_{j,t}}$, estimates the advantage, and updates ${{\theta }_{j}}$ and ${{w}_{j}}$;
			\EndIf
	    	\Until{\text{Stopping criteria}}			
	\end{algorithmic}}
\end{algorithm}

\section{Simulation Results}
\label{SecSimResults}
In the simulations, the entire frequency band is divided into 8 frequency domain channels.
The frequency hopping schedule of the malicious sensing device $r$ and the user $u$ can be formulated as $f\left( t \right)=k\cdot t+b$. Specifically, the frequency utilization of the malicious sensing device sweeps from the 1\,st to the 8\,th frequency domain channel with $k=1$ and $b=0$. While the frequency utilization of the user sweeps from the 8\,th to the 1\,st frequency domain channel with $k=-1$ and $b=8$. The values of the immediate reward are set s ${{R}_{1}}=1$, ${{R}_{2}}=0$, ${{R}_{3}}=-5$ and ${{R}_{4}}=5$.

In terms of changes in the mode of operation, the transition from the searching mode to the lock-on mode means an increasing threat level.
In the simulations, we evaluate the jamming success rate over confrontations between the jammer and the malicious sensing device, where a confrontation occurs at a critical instant when the jamming fails or succeed to make the sensing device remain operating in the identical mode or switch to a mode with lower threat level.
During the time when the sensing device continuously operates in a mode, confrontation can occur several times.
For instance, if the sensing device, which operates in the searching mode, has detected the target in the first two detections, the third detection is regarded as a critical instant, where a confrontation happens. In the third detection, if the noise jamming interfere with the sensing device and the target is not detected, the fourth detection would be considered as another critical instant, when the second confrontation occurs; otherwise, the malicious sensing device switches to the tracking mode.
Let $N_{m,0}$ denote the total number of confrontations in the $m$\,th occurrence of a certain mode of operation, and let ${{N}_{m,j}}$ represent the number of times the jammer succeeds in the confrontation in the total $N_{m,0}$ confrontations. Hence, the empirical jamming success rate measured in the $m$\,th occurrence of a certain mode of operation can be defined as
\begin{equation}
\begin{split}
{{P}_{m,j}}=\frac{{{N}_{m,j}}}{{{N}_{m,0}}}\times 100\%\,.
\end{split}
\end{equation}

\begin{figure}[H]
\centering
\includegraphics[width = 3.2in]{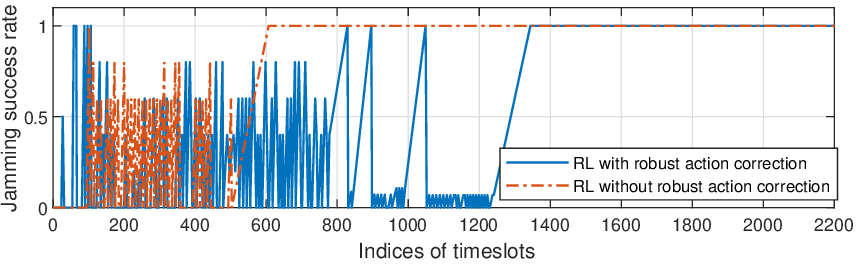}
\caption{Comparison of jamming success rate in the offline training phase.}
\label{FCRO}
\end{figure}
\begin{figure}[H]
	\centering
	\includegraphics[width = 3.2in]{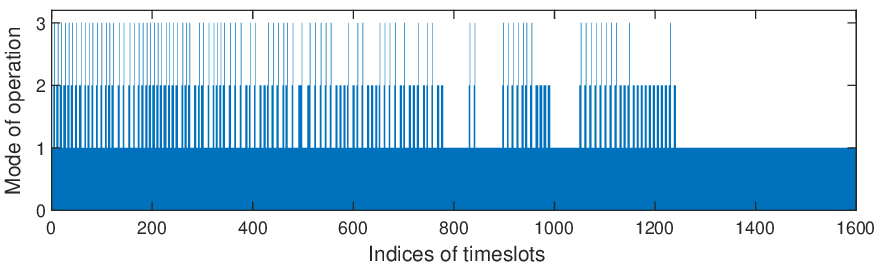}
	\caption{Mode of operation of malicious sensing device in the offline training phase. 1 indicates that the malicious device is working in the searching mode; 2 indicates tracking mode; 3 indicates lock-on mode.}
	\label{MO}
\end{figure}
\vspace{-0.25cm}
\begin{figure}[H]
	\centering
	\includegraphics[width = 3.2in]{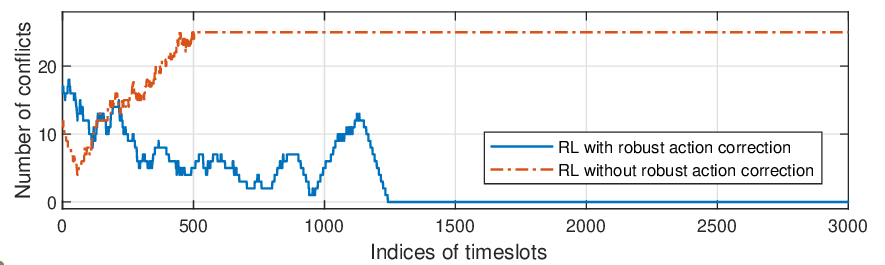}
	\caption{Comparison of the number of conflicts in the offline training phase.}
	\label{CNO}
\end{figure}
Fig. \ref{FCRO} invsestigates jamming success rates achieved by the AC method-based frequency selection algorithms with/without robust action correction in the offline training phase. It can be seen from Fig. \ref{FCRO} that jamming success rates achieved by the AC method-based algorithm finally staturates, reaching a success rate of 100\%. In conjunction with Fig. \ref{MO}, which illustrates the mode of operation of malicious sensing device transition, it can be inferred that the malicious sensing device continuously gets stuck in searching mode and poses no threat to the target. Fig. \ref{FCRO} also illustrates that the AC method-based algorithm with robust action correction may converge slower than that without robust action correction due to the action correction processes in the training phase.

Fig. \ref{CNO} depicts the number of frequency-domain conflicts as a function of timeslots. It is shown that the number of conflicts achieved by the AC method-based frequency selection algorithm without robust action correction scales with the timeslots and eventually saturates, reaching a value of 25. In contrast, the AC method-based frequency selection algorithm with robust action correction indicates a gradual decrease in the number of conflicts,, eventually reducing to zero. Such an observation indicates that the robust action correction process can effectively mitigate frequency-domain conflicts between frequencies for noise jamming and those for non-cooperative uplink transmissions.

\begin{figure}[H]
	\centering
	\includegraphics[width = 3.2in]{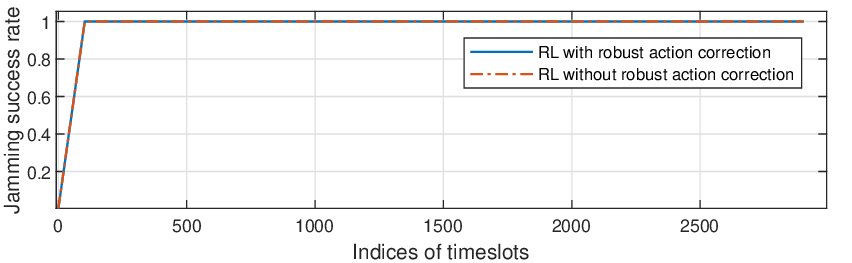}
	\caption{Comparison of jamming success rate in the online inference phase.}
	\label{z1}
\end{figure}
\begin{figure}[H]
	\centering
	\includegraphics[width = 3.2in]{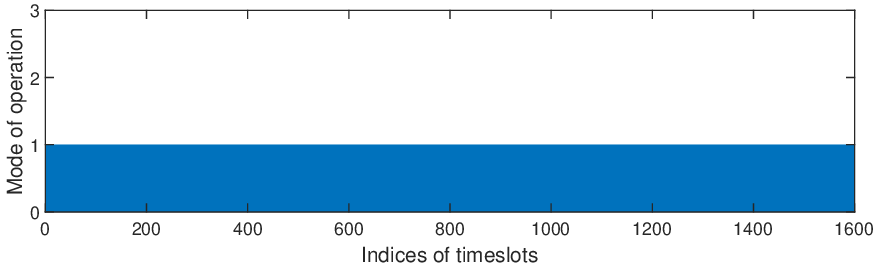}
	\caption{Mode of operation of malicious sensing device in the online inference phase. 1 indicates that the malicious device is working in the searching mode; 2 indicates tracking mode; 3 indicates lock-on mode.}
	\label{z2}
\end{figure}
\begin{figure}[H]
	\centering
	\includegraphics[width = 3.2in]{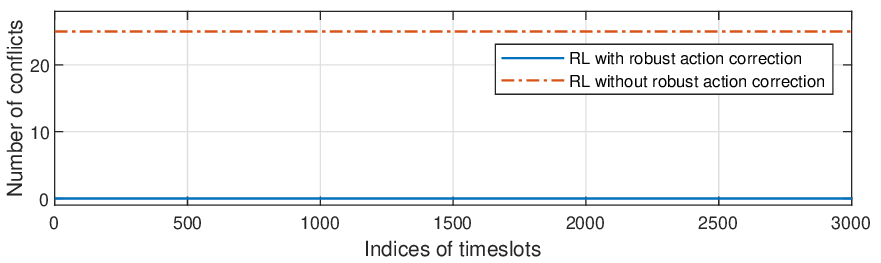}
	\caption{Comparison of the number of conflicts in the online inference phase.}
	\label{z3}
\end{figure}
It can be observed from Fig. \ref{z1} that both the AC method-based frequency selection algorithms with/without robust action correction can achieve a jamming success rate of 100\% in the online inference phase. In the meanwhile, Fig. \ref{z2} illustrates that the malicious sensing device continuously gets stuck in searching mode during the online inference phase. Fig. \ref{z3} demonstrates that in the online inference phase, the AC method-based frequency selection algorithm without robust action correction always suffers from 25 conflicts in each timeslot, whereas the algorithm with robust action correction encounters at zero conflict. The above observations imply that given a proper training set, the AC method-based frequency selection algorithm with robust action correction can configure frequency-domain resources that can make noise jamming effectively interfere with the malicious sensing device and simultaneously avoid interference with non-cooperative uplink transmissions.

\section{Conclusions}
\label{SecConclusions}
This paper has investigated the physical layer security issues for 6G mobile communication systems, in terms of preventing malicious sensing by jamming. To this end, we have proposed a novel AC method-based frequency selection scheme for noise jamming, integrated it with a robust action correction method, which aims at avoiding potential frequency-domain conflicts between jamming and non-cooperative but friendly uplink transmissions. Numerical results illustrate that by integrating the robust action correction scheme, the proposed AC method-based frequency selection algorithm can significantly reduce the conflicts compared to that without robust action correction. Benefiting from this, the proposed method can guarantee minimal interference to the non-cooperative user during jamming the malicious sensing device.


\begin{thebibliography}{99}
	\bibitem{MAHG23}M. Mitev, A. Chorti, H. V. Poor and G. P. Fettweis, "What Physical Layer Security Can Do for 6G Security," in IEEE Open Journal of Vehicular Technology, vol. 4, pp. 375-388, 2023.
	\bibitem{QQ23}Q. Chen and Q. Du, "Differentially Pre-coded Polar Codes for Physical Layer Security," 2023 IEEE 98th Vehicular Technology Conference (VTC2023-Fall), Hong Kong, 2023, pp. 1-6.
	\bibitem{BKBYYH23}B. Yang, K. Li, B. Jiu, Y. Zhao, Y. Wang and H. Liu, "An Intelligent Jamming Strategy Design Method Against Frequency Agility Radar," 2023 IEEE International Radar Conference (RADAR), Sydney, Australia, 2023, pp. 1-6.
	\bibitem{ZYYGWG22}Z. Feng, Y. Xu, Y. Jiao, G. Li, W. Li and G. Fang, "Fight Against Smart Communication Rival: An Intelligent Jamming Approach With Trend-Oriented Efficacy Evaluation," in IEEE Wireless Communications Letters, vol. 11, no. 11, pp. 2290-2294, 2022.
	\bibitem{GYWJYH23}G. Xu, Y. Zhang, W. Huo, J. Pei, Y. Zhang and H. Yang, "A Cognitive Jamming Decision-making Method for Multi-functional Radar Based on Threat Assessment," 2023 IEEE Radar Conference (RadarConf23), San Antonio, TX, USA, 2023, pp. 1-6.
	\bibitem{ZYSY23}Z. Pan, Y. Li, S. Wang and Y. Li, "Joint Optimization of Jamming Type Selection and Power Control for Countering Multifunction Radar Based on Deep Reinforcement Learning," in IEEE Transactions on Aerospace and Electronic Systems, vol. 59, no. 4, pp. 4651-4665, 2023.
	\bibitem{YLSTJ23}Y. Fang, L. Zhang, S. Wei, T. Wang and J. Wu, "Online Frequency-Agile Strategy for Radar Detection Based on Constrained Combinatorial Nonstationary Bandit," in IEEE Transactions on Aerospace and Electronic Systems, vol. 59, no. 2, pp. 1693-1706, 2023.
	\bibitem{MSDAA23}M. M. H. Qazzaz, S. A. Zaidi, D. McLernon, A. Salama and A. A. Al-Hameed, "Low Complexity Online RL Enabled UAV Trajectory Planning Considering Connectivity and Obstacle Avoidance Constraints," 2023 IEEE International Black Sea Conference on Communications and Networking (BlackSeaCom), Istanbul, Turkiye, 2023, pp. 82-89.
	\bibitem{N23}N. Geng et al., "A Reinforcement Learning Framework for Vehicular Network Routing Under Peak and Average Constraints," in IEEE Transactions on Vehicular Technology, vol. 72, no. 5, pp. 6753-6764, 2023.
	\bibitem{YCYF21}Y. Huang, C. Hao, Y. Mao and F. Zhou, "Dynamic Resource Configuration for Low-Power IoT Networks: A Multi-Objective Reinforcement Learning Method," in IEEE Communications Letters, vol. 25, no. 7, pp. 2285-2289, 2021.
	\bibitem{RA18}R. S. Sutton and A. G. Barto, Reinforcement Learning: An Introduction,
	2nd ed. Cambridge, MA, USA: MIT Press, 2018.
	\bibitem{YXQ23}Y. Huang, X. Zhu and Q. Wu, "Intelligent Spectrum Anti-Jamming With Cognitive Software-Defined Architecture," in IEEE Systems Journal, vol. 17, no. 2, pp. 2686-2697, 2023.
	\bibitem{GKMTCY18}G. Dalal, K. Dvijotham, M. Vecerik, T. Hester, C. Paduraru, Y. Tassa, "Safe Exploration in Continuous Action Spaces," arXiv:1801.08757 [cs], 2018.
\end{thebibliography}
\end{document}